# Investigations of Surface Quality and SRF Cavity Performance

G. Wu[1], M. Ge[1], P. Kneisel[2], K. Zhao[3], L. Cooley[1], J. Ozelis[1], D. Sergatskov[1] and C. Cooper[1],
[1]*Fermilab,* [2]*Jefferson Lab,* [3]*Peking University*

*Abstract*—Magnetic field enhancement has been studied in the past through replica and cavity cutting. Considerable progress of niobium cavity manufacturing and processing has been made since then. Wide variety of single cell cavities has been analyzed through replica technique. Their RF performances were compared in corresponding to geometric RF surface quality. It is concluded that the surface roughness affects cavity performance mostly in secondary role. The other factors must have played primary role in cavity performance limitations.

## I. Introduction

RECENT cavity processing statistics indicate that the development of RF superconductivity has reached a stage where more and more cavities were limited by quench and not by field emissions [1, 2]. Among those quench limiting cavities, more than half of them were limited by identifiable surface features, namely pits or bumps. The quench field ranges from 12.7 MV/m up to 42 MV/m.

The high resolution optical inspection tools presently in use have captured many pictures of cavity surface defects [3, 4]. Many of them caused cavity to quench, as corroborated by temperature mapping systems (T-map) [5] or second sound detection systems [6]. The optical image offers little help in resolving the fine details of the surface defects. Replica techniques were used to better understand cavity inner surface features [7]. They have recently been used to replicate the cavity inner surfaces of ILC 9-cell cavities including equator and iris regions [8].

Although the replica technique does not capture geometric details below 1 micron, we believe the amount of information is quite sufficient for relative comparisons, even detailed enough to calculate the magnetic field enhancement at these features. In the case of the magnetic field enhancement at the cavity surface, atomic scale information of the surface is not necessary. This was demonstrated by the analysis of large grain 1-cell cavity surfaces.

Other than the geometric defects inside the cavity, general surface roughness is broadly characterized through cavity surface replicas.

## II. Method of Surface Characterization

Many cavities were cold tested with thermometers attached to the high magnetic field region. Most of the cavities were inspected using an optical telescope [9]. Replicas were made to selected equator regions based on optical inspection results [8]. Cavities being studied were mostly electron beam welded. We also studied cavities being made through hydroforming and spinning. Processing included Buffer Chemical Polishing (BCP), Electropolishing (EP), Barrel Polishing (Tumbling) and their combinations.

All the replicas were scanned using a high resolution profilometer (KLA Tencor Model P-16 with 0.8 µm diameter L-stylus) to obtain geometric contours. s. The profilometer has a horizontal resolution below 1 µm and vertical resolution ~1 Å [10].

Geometric data were categorized in two classes. One class is for special features. The specific features were analyzed case by case to closely match the cavity performances. The other class is for general areas with no specific geometric feature. For the general areas, roughness data Rz is given for three typical regions in the cavity such as weld seam, heat affected zone and rest of the cavity area. The Rz data for each region is available for three different surface sizes (1mm x 1mm, 250 µm x 250 µm and 50 µm x 50 µm).

A simple surface model, as shown in Fig. 1, and similar to the previous magnetic field enhancement models, was used [11, 12].

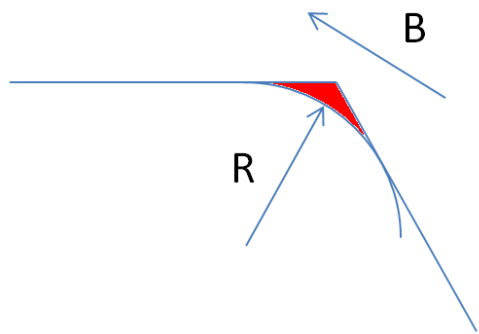

Fig. 1: A simple surface model.

Manuscript received 3 August 2010. This work was supported in part by the U.S. Department of Energy under contract number DE-AC02-07CH11359.
G. Wu, M. Ge, L. Cooley, C. Cooper, J. Ozelis and D. Sergatskov are with Fermilab, Batavia, IL 60510, USA. (G. Wu: 630-840-5409; e-mail: genfa@fnal.gov).
P. Kneisel is with Jefferson Lab, Newport News, VA, 23606, USA.
K. Zhao is with Peking University, Beijing, China.

As one would expect when magnetic field at the edge of the defect exceeds the thermal critical magnetic field, the magnetic flux then enters the corner area deeper than the penetration depth, depending on the field enhancement factor. At this time, the defect can be divided into one lossy corner and a remaining flux-free body. According to previous field enhancement models, such lossy corners remain small in both depth and length scale. Such lossy corners could be harmful when present in a wide area such as in cavities that were BCP processed. They cause very limited harm in the case of large grain cavities or EP cavities with isolated defects. Such analysis including detailed thermal analysis will be published in a future paper.

The above described model confirmed that the loss of resolution below 1 μm of replica is indeed not critical for the investigation of cavity RF surfaces.

### III. SURFACE DEFECTS

While generic surface quality such as equator weld and surface roughness do affect cavity performance, they only play such a role in a collective fashion, such as the case of fine grain cavities being BCP processed. For cavities that are made from large grain or are EP processed, the geometric quality of welds, grain boundaries, and surface roughness plays only a secondary role. A large number of cavity defects have been characterized and compared to their cavity performances. Several typical examples are described in this section.

Nine-cell cavity TB9ACC017 was tested with a partial T-map which had several temperature sensors located on the equator and near a quench location. The cavity quenched at a rather low gradient of 12.3 MV/m, with the optically-detected defect location registering the highest temperature rise during a quench. Fig. 2 shows this defect within the equator weld.

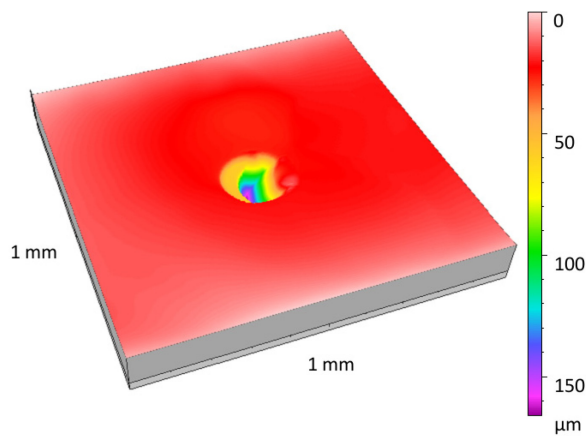

Fig. 2: Equator defect in cavity TB9ACC017.

Nine-cell cavity AES001 was tested with a similar partial T-map [5] and geometric defects of twin bumps presumably caused the cavity to quench at 21.8 MV/m even after repeated EP processing. This feature is shown in the replica images in Fig. 3.

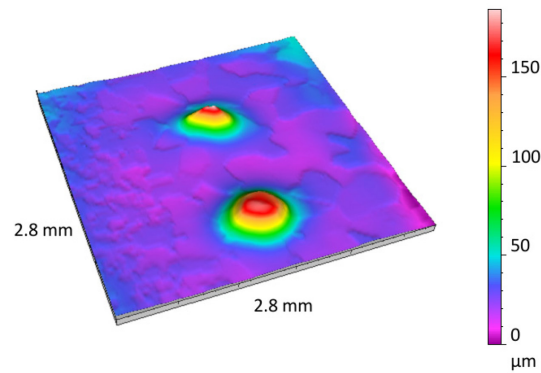

Fig. 3: Equator twin bumps in cavity AES001.

Single-cell cavity TE1AES004 was tested with a full body temperature mapping system (T-map) at Jefferson Lab [13]. A geometric defect was found in the equator weld as shown in Fig. 4.

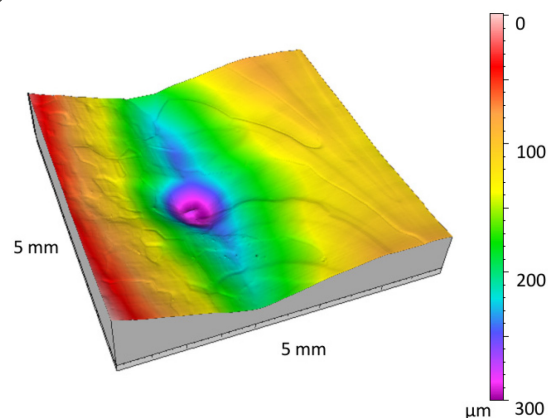

Fig. 4: Equator defect in cavity TE1AES004.

This cavity quenched at 39 MV/m with a peak magnetic field of 168 mT. T-map results indicated the quench was elsewhere and this defect location registered no elevated heating.

Single-cell cavity PKU-LG1 reached 43 MV/m when tested at JLAB. Its internal surface was studied using an optical camera and an area of interest was selected for generation of a replica. Fig. 5 shows one of the very sharp grain boundaries near the equator.

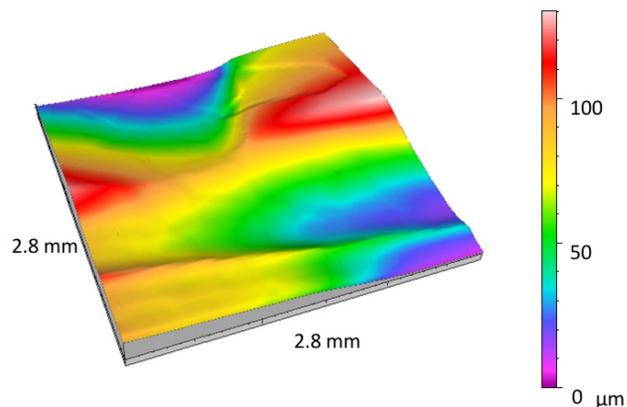

Fig. 5: One equator grain boundary of 1-cell cavity PKU-LG1.

Table 1 lists the geometrically derived field enhancement factors for the three defects discussed above. The field enhancement factor was calculated using a 2D model.

TABLE 1 FIELD ENHANCEMENT FACTOR OF A CAVITY DEFECTS

| Cavity | Field enhancement factor | Maximum H field [mT] |
|---|---|---|
| TB9ACC017 | 2.2 | 54.1 |
| AES001 | 1.5 | 96.8 |
| TE1AES004 | 1.2 | 168.0 |
| PKU-LG1 | 1.6 | 185.0 |

Cavities listed in Table 1 were EP processed except the large grain cavity (PKU-LG1) which received BCP processing. The maximum local magnetic field can be obtained by multiplying the calculated field enhancement factor and the measured cavity maximum H field. For example, the local maximum magnetic field near the defect in TB9ACC017 was much lower than the thermal critical magnetic field of niobium. For the case of PKU-LG1, the local magnetic field at the grain boundary apparently far exceeded the thermal critical magnetic field of niobium. From studies based on these 4 defect-containing cavities it appears that the geometric topology serves only a secondary role in limiting cavity performances. Intrinsic material imperfections remain the primary cause for poor performance in SRF cavities. Geometric imperfections can be mitigated by using the standard EP process for cavities to achieve high performance when material imperfections were absent.

## IV. ROUGHNESS

For cavities that have no apparent geometric defects or with geometric imperfections which are shown to cause little degradation in cavity performance, the surface roughness was obtained in areas of 1 mm$^2$. The roughness data is taken at area that is not in the heated affected zone. For large grain or single grain cavities, the target area is within a single grain.

The BCP processing was standard cavity process where cavity material was removed in the amount of 120 μm in thickness. In the case of large grain, the amount of material removal was at 300 μm in thickness. EP was in standard horizontal rotation fashion. Light EP represents cavity material removal around 40 μm in thickness, while heavy EP removes the material greater than 120 μm in thickness. Tumbling is a combination of mechanical barrel polishing (120 μm) and light EP (40 μm).

Table 2 lists the surface roughness for those typical cavities, their processing method and their highest peak magnetic field. Cavities are all single cell cavities with TESLA shape. Each cavity can be considered to be representative of many cavities similarly processed, with similar replica results. The ratio of maximum magnetic field to accelerating gradient is 4.3 and 4.26, corresponding to the TESLA end cell or center cell shape.

TABLE 2 GENERAL AREA SURFACE ROUGHNESS

| Cavity | Processing | Ra (μm) | Rz (μm) | Hmax [mT] |
|---|---|---|---|---|
| NR-6 | BCP | 0.257 | 1.210 | 115 |
| PKU-SC1 | BCP | 0.208 | 1.004 | 128 |
| TE1ACC005 | Light EP | 0.116 | 0.438 | 163 |
| PKU-LG1 | BCP | 0.0606 | 0.306 | 183 |
| TE1ACC003 | Heavy EP | 0.072 | 0.303 | 173 |
| TE1ACC004 | Tumbling and light EP | 0.049 | 0.201 | 174 |

Fig. 6 shows the plot of the surface roughness (Rz) and the maximum magnetic field cavities have achieved.

The cavities listed in Table 2 and Fig. 6 are very good examples representing a large number of cavities divided into six major categories that we have studied.

All cavities were made of high quality niobium with high residual resistivity ratios between 300 and 500.

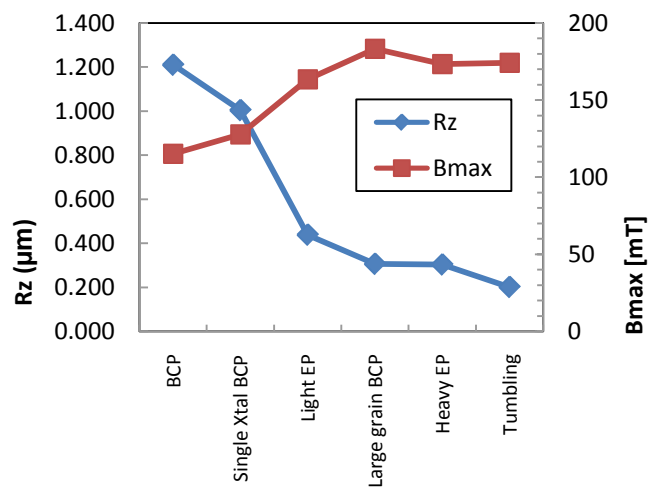

Fig. 6: The cavity surface roughness compared to the maximum surface magnetic field.

Cavity performance becomes less dependent on the surface roughness as the maximum magnetic field reached a plateau, even while the surface was continuously improved as with heavy EP or tumble polishing. This again indicates that the geometric quality of a cavity's RF surface plays a secondary role, similar to the situation when apparent geometric defects were present.

## V. CONCLUSIONS

Internal RF surfaces of a large number of cavities have been analyzed. These cavities represented six major categories of modern SRF cavity technology (seamless cavities made by spinning or hydroforming are not discussed in detail here). As we have observed, an equator weld defect only affects cavity performance when there is a concurrent apparent material defects.

The defects can be divided as geometric imperfections or intrinsic to the material. Geometric imperfections played a secondary role in limiting cavity performance. Among those geometric imperfections, the apparent defects are not necessarily harmful other than bearing the risk of trapping acidic water during processing. Standard processing can easily reduce the magnetic field enhancement and push the cavities to a higher gradient in cavities that have no apparent material or geometric defects; the surface roughness is sufficiently good for both fine grain cavities with the standard EP process and for large grain cavities using the standard BCP process.

ACKNOWLEDGMENT

We would like to thank the support effort provided by A. Rowe, D. Bice, C. Baker, B. Stone, M. Carter, D. Marks, D Burk and D. Hicks. We also thank G. Ciovati, M. Morrone for one of the cavity tests and useful discussions. K. Swanson provided editorial help.